\documentclass[prc,aps,prX,twocolumn,groupedaddress,nofootinbib]{revtex4-1}
\usepackage{amsmath}
\usepackage{amsfonts}
\usepackage{amssymb}
\usepackage[english]{babel}
\usepackage{graphicx}
\usepackage{epstopdf}
\usepackage{hyperref}
\usepackage{color}
\usepackage{dcolumn}% Align table columns on decimal point
\usepackage{bm}% bold math
\everymath{\displaystyle}

\begin{document}

\title{Parity-odd effects in heavy-ion collisions in the HSD model}

\author{\firstname{Oleg}~\surname{Teryaev}}
\email{teryaev@theor.jinr.ru}
\affiliation{Joint Institute for Nuclear Research, 141980 Dubna
(Moscow region), Russia}
\affiliation{Dubna International University, Dubna (Moscow region) 141980,
Russia}
\author{\firstname{Rahim}~\surname{Usubov}}
\email{usubov@theor.jinr.ru}
\affiliation{Joint Institute for Nuclear Research, 141980 Dubna
(Moscow region), Russia}
\affiliation{Moscow Institute of Physics and Technology, Dologoprudny
  (Moscow region) 141700, Russia}
\date{\today}

\begin{abstract}

Helicity separation effect in non-central heavy ion collisions is
investigated using the Hadron-String Dynamics (HSD) model.
Computer simulations are done to calculate velocity and hydrodynamic
helicity on a mesh in a small volume around the center of the reaction.
The time dependence of hydrodynamic helicity is observed for
various impact parameters and different calculation methods.
Comparison with a similar earlier work is carried out. A new quantity related 
to jet handedness is used
to ananlyze particles in the final state. It is used to probe for p-odd
effects in the final state. 

\end{abstract}

\maketitle

\section{Introduction}

C(P) odd effects in heavy ion collisions are under intensive theoretical
investigation nowadays \cite{intro}. C(P) odd effects can manifest themselves in
several ways.  Chiral Magnetic Effect that leads to appearance of
electric current in the presence of external magnetic field. Another example
is Chiral Vortical Effect. It is caused by the presence of vorticity in the QCD
matter. The most interesting effects are resulted by the presence of vorticity
developing in non central heavy ion collisions which may lead
\cite{Rogachevsky:2010ys} to neutron asymmetries. 

Thus it is important to find out if vorticity really exists in different
models and calculate related quantities such as helicity to observe their
time evolution.

Vorticity and helicity in heavy ion collisions were investigated
in \cite{helsep0,cshid}.

Classic vorticity as well as its relativistic generalization were
calculated in 3+1 dimensional hydrodynamic model. Velocity circulation has
also been analyzed in \cite{cshid}.

Another interesting quantity that can be studied is helicity:
$$ H = \int (\vec{v}, rot\vec{v})dV. $$
It can be divided into two parts depending on the $v_y$ component of velocity:
$$ H_{\uparrow} = \int (\vec{v}, rot\vec{v})dV,   v_y>0 $$ and
$$ H_{\downarrow} = \int (\vec{v}, rot\vec{v})dV,   v_y<0. $$
If there is non zero medium vorticity with a dominating direction, these
two quantities will have different signs. Time dependence of these quantities
can give additional information on medium vorticity.
Helicity has been studied in \cite{helsep0} with special emphasis on its
time dependence. It was shown that helicity separation can be observed in
QGSM model. The time dependence of other integral quantities regarding
helicity and vorticity were also calculated in that work.

The aim of current paper is to investigate vorticity and helicity in 
heavy ion collisions with the help of HSD model \cite{hsdref}.
%The HSD model is based on the BUU model.
The HSD modelling program provides the numerical solution of a set of
relativistic transport equations for particles with in-medium selfenergies.
Comparison to the similar results obtained in a QGSM model \cite{helsep0}
is also made.

We also study the directly observable quantity - handedness. It is a
modification of the variables proposed in \cite{emth}, \cite{nachtm}
to study particle polarisation.\\

\section{Modelling velocity field}

The heavy ion collision modelling was done with the help of slightly modified
HSD program \cite{hsdref}. Au + Au collisions with different impact parameters
and with bombarding energy of 12.38 GeV per nucleon were simulated, which
corresponds to $\sqrt{s} = 5 GeV$ in the center-of-mass frame.
Before collision
nuclei travel along $Z$ axis. Distance between centers of the nuclei is $b$
along the $X$ axis. Plane $y = 0$ is called reaction plane.

Velocity field was calculated using energies and momenta of all particles
in the final state. All final state quantities are given in the center of mass
frame, calculations are carried out in the same frame of reference. The space
was represented with a three dimensional mesh. Each cell is a rectangular cuboid
the following parameters: $\Delta x = \Delta y = \gamma \Delta z = 0.6 fm$,
where $\gamma$ is the gamma factor of the center of mass frame. Velocity
field was computed as follows:
$$ \vec{v}(x, y, z) = \frac{\sum_i\sum_j \vec{P}_{ij}(x, y, z)}
{\sum_i\sum_j E_{ij}},$$
where, $i$ represents the number of the event and $j$ represents the number of
the particle. Cells with at least two particles were taken into account.

Velocity field obtained this way was used to calculate helicity ($H$) and
vorticity ($rot\vec{v}$) All results presented were calculated
using a basic two point difference formula for derivatives. As we will see
later, a more sophisticated formula for derivatives~\eqref{eq:newderiv1},
~\eqref{eq:newderiv2},~\eqref{eq:newderiv3} doesn't give better
results. For vorticity distribution weighted $(rot\vec{v})_y$ was used, as
suggested in \cite{cshid}.
Vorticity in each cell $(m, n, k)$ was weighted by the factor of 
$w_{m, n, k}$:
$$w_{m, n, k} = \frac{E_{m, n, k}N_{cells}}{2E_{total}},$$
where $E_{m, n, k}$ is the sum of energies of all particles in the cell
$(m, n, k)$, $N_{cells}$ is the total number of cells and $E_{total}$ is
the total energy in the whole volume. This weighted vorticity was averaged
over all $x-z$ layers to get a single $x-z$ layer at different times.
In order to observe
helicity separation cells are divided into two groups depending
on sign of velocity component $v_y$ normal to the reaction plane. $H$ was
calculated for both kinds of cells separately.\\
\section{Results and comparison}
\subsection{Weighted Vorticity}

\begin{figure}
  \centering
  \includegraphics[width=\hsize]{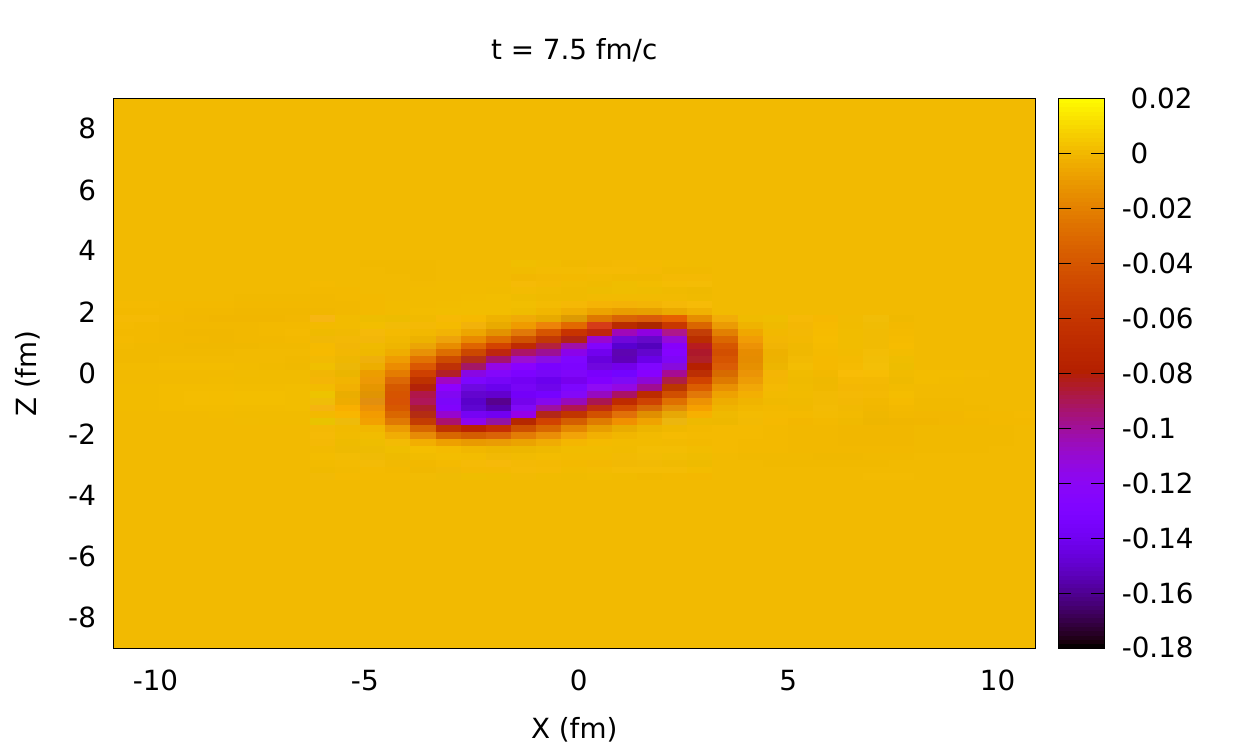}
  \caption{Weighted $y$-component of the vorticity ($c/fm$) averaged over
    all $x-z$ layers at $7.5 fm/c$, impact parameter $b=8fm$.
    Average value is $-4.4395\cdot 10^{-2} c/fm$.}
  \label{fig:vortdist7}
\end{figure}

\begin{figure}
  \centering
  \includegraphics[width=\hsize]{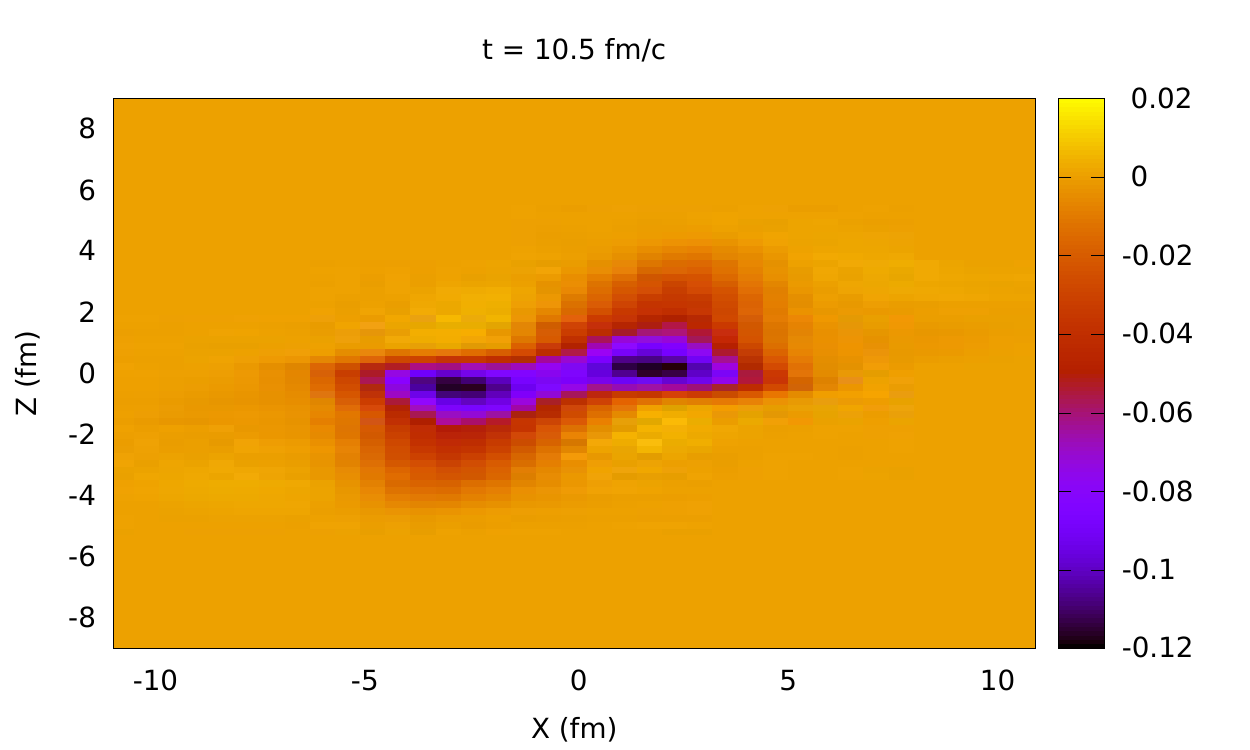}
  \caption{Weighted $y$-component of the vorticity ($c/fm$) averaged over
    all $x-z$ layers at $10.5 fm/c$, impact parameter $b=8fm$.
    Average value is $-2.2800\cdot 10^{-2} c/fm$.}
  \label{fig:vortdist10}
\end{figure}

\begin{figure}
  \centering
  \includegraphics[width=\hsize]{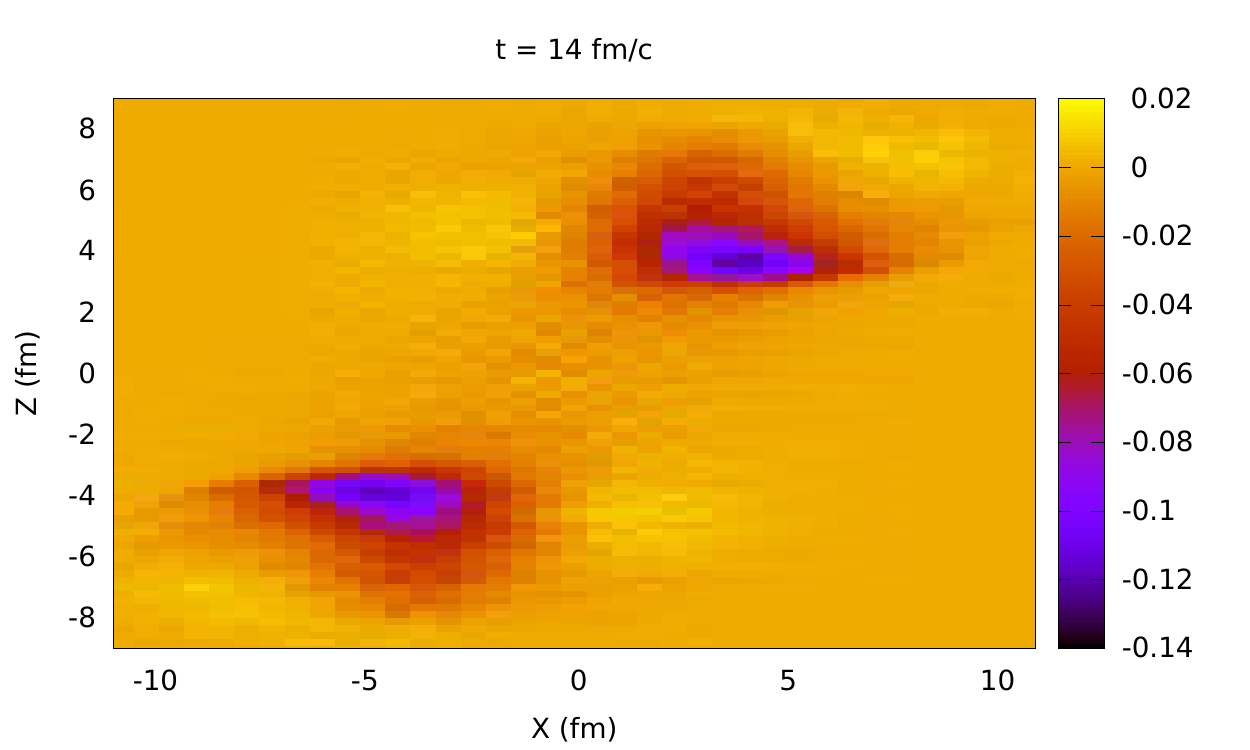}
  \caption{Weighted $y$-component of the vorticity ($c/fm$) averaged over
    all $x-z$ layers at $14 fm/c$, impact parameter $b=8fm$.
    Average value is $-1.2452\cdot 10^{-2} c/fm$.}
  \label{fig:vortdist14}
\end{figure}

\begin{figure}
  \centering
  \includegraphics[width=\hsize]{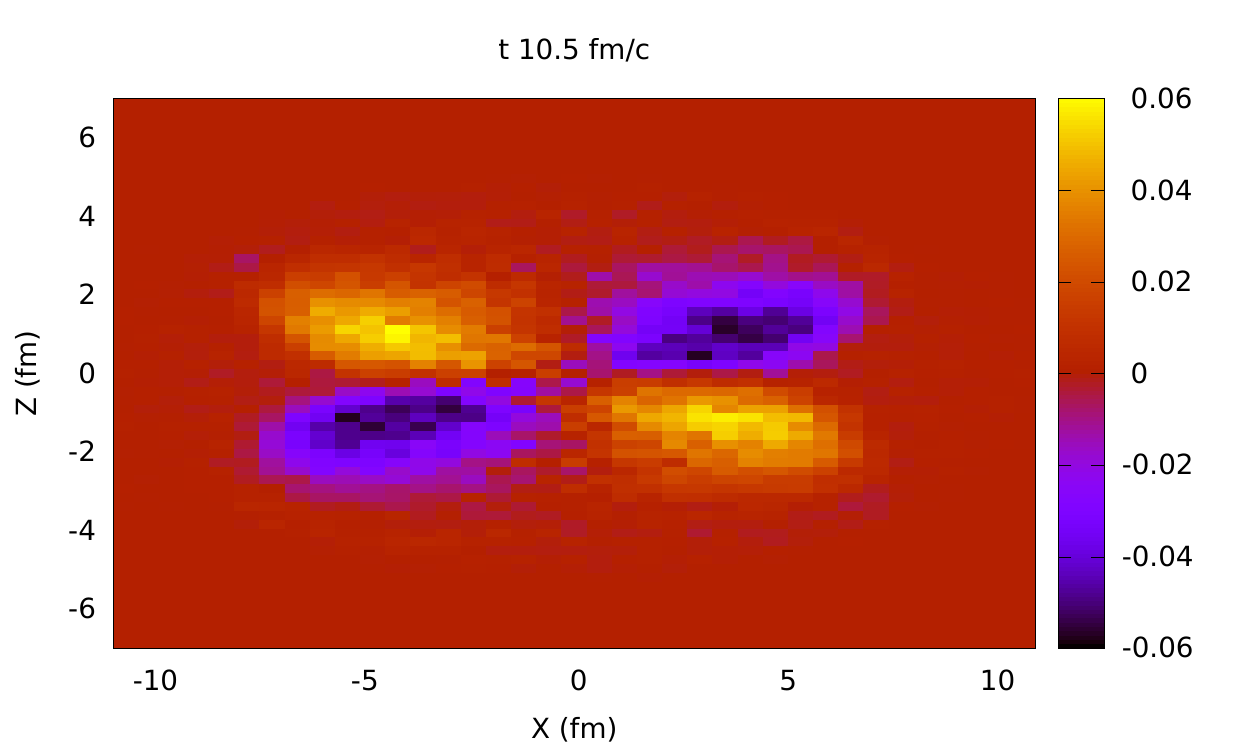}
  \caption{Weighted $y$-component of the vorticity ($c/fm$) averaged over
    all $x-z$ layers at $10.5 fm/c$, impact parameter $b=0fm$.
    Average value is $-1.2\cdot 10^{-5} c/fm$.}
  \label{fig:vortdistb010}
\end{figure}

In this subsection we present the weighted $y$-component of vorticity
averaged over all $x-z$ layers at different times. To observe time evolution
of weighted vorticity it was plotted at three different time moments:
7.5 $fm/c$, 10.5 $fm/c$ and 14 $fm/c$, impact parameter $b=8fm$
(Figures~\ref{fig:vortdist7},~\ref{fig:vortdist10},~\ref{fig:vortdist14}).
A similar plot for $t=10.5 fm/c$ and impact parameter $b=0fm$ is included
(Figure~\ref{fig:vortdistb010}). In the latter case weighted $y$-component
of the vorticity is less in magnitude and there are regions with positive
as well as negative $y$-component of the vorticity. The overall average
is decreasing in time. As for the $b=0fm$ case
(Figure~\ref{fig:vortdistb010}), we notice that the
average value of weighted $y$-component of vorticity is negligible with
relation to the same time moment $t=10.5fm/c$ with impact parameter $b=8fm$
(Figure~\ref{fig:vortdist10}).

\subsection{Helicity}

%[h!]
\begin{figure}
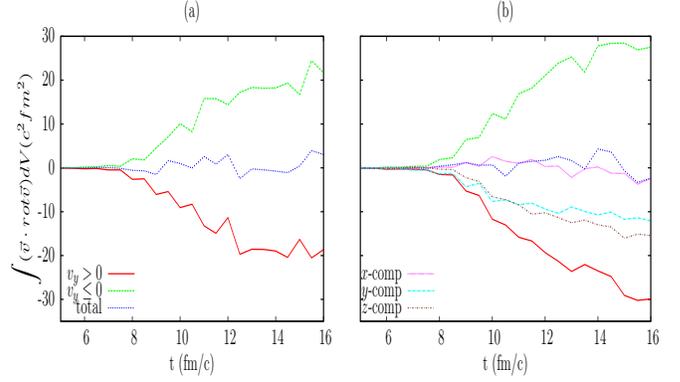

  \centering
  {\fontsize{15}{19} \selectfont
  \resizebox{\hsize}{5.2cm}{\input{plots/plotb4.tex}
    \input{plots/plotb8.tex}}
  }
  %\resizebox{0.4\hsize}{!}{\input{plots/plotb8.tex}}
  \caption{ Helicity: $H_{\uparrow}$ and $H_{\downarrow}$, impact parameter
    $b = 4 fm/c$ (a) and $b = 8 fm/c$ (b).}
  \label{fig:helcomp}
\end{figure}

Main results obtained in HSD model for helicity separation are presented in 
Figure ~\ref{fig:helcomp}. Simulations in HSD model manifest helicity
separation similar to the separation in QGSM model. Along with this there is a
notable shift in time (about $6 - 7 fm/c$). Helicity separation begins
later than in \cite{helsep0}. This may be explained by the difference in
initial state of the nuclei. In the HSD simulation program the nuclei
are initial at distance $~7 fm$ apart from each other. Since they start
off at a significant distance it takes some time for them to collide
\footnote{Authors thank M. Baznat for this valuable observation.}.

Magnitudes of $H_{\uparrow}$ and $H_{\downarrow}$ are also different from the same
quantities in \cite{helsep0}. $H_{\uparrow}$ subdivided by components
in scalar product is shown in Figures ~\ref{fig:helcomp}.
In both models the $x$ component doesn't give a significant contribution in
$H$. For the QGSM model there is a difference between $y$ and $z$ component
contributions. However, there is no such tendency for the HSD model: both
components give contribution of similar magnitude.
%[h!]
\begin{figure}[!t]
  \centering
  \resizebox{\hsize}{5cm}{\input{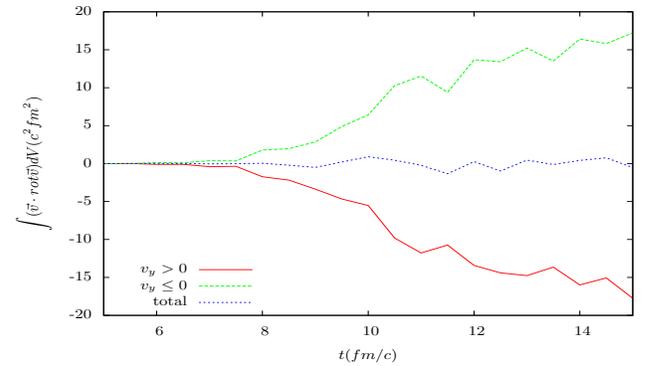}}
  \caption{Results for new derivative formulas parameter $b = 4 fm/c$.}
  \label{fig:newderiv}
\end{figure}

The same quantity was calculated using another formula for velocity derivatives
in helicity for comparison. It can be interesting to see if a more
accurate formula can improve the result. Let us calculate derivatives as 
follows:
\begin{multline}\label{eq:newderiv1}
  \partial_x v_{\alpha}(m,n,k) = \\
  \frac{1}{8h_x}\sum_{i=-1, 1} \sum_{j=-1, 1}
  \{v_{\alpha}(m + 1, n + i, k + j)\\
  - v_{\alpha}(m - 1, n + i, k + j)\},
\end{multline}
\begin{multline}\label{eq:newderiv2}
  \partial_y v_{\alpha}(m,n,k) = \\
  \frac{1}{8h_y}\sum_{i=-1, 1} \sum_{j=-1, 1}
  \{v_{\alpha}(m + i, n + 1, k + j)\\
  - v_{\alpha}(m + i, n - 1, k + j)\},
\end{multline}
\begin{multline}\label{eq:newderiv3}
  \partial_z v_{\alpha}(m,n,k) = \\
  \frac{1}{8h_z}\sum_{i=-1, 1} \sum_{j=-1, 1}
  \{v_{\alpha}(m + i, n + j, k + 1)\\
  - v_{\alpha}(m + i, n + j, k - 1)\},
\end{multline}

where $h_x, h_y, h_z$ are the cell sizes along $x, y$ and $z$ axis respectively,
$m, n, k$ are discrete coordinates on the mesh. This method of calculation
uses higher order discrete derivative and averaging over four derivatives
calculated at different points.
The new formula for derivatives, however, doesn't give any new or improved
results (Figure ~\ref{fig:newderiv}).
%% \begin{figure}[!b]
%% %  \centering
%%   \includegraphics[width=\hsize, height=8cm]{./otnv5_oct_2013}\\
%%   \caption{Integral values calculated for impact parameter $b=8fm$, in QGSM
%%     model \cite{helsep0}.}
%%   \label{fig:integrqgsm}
%% \end{figure}
\begin{figure}[!h]
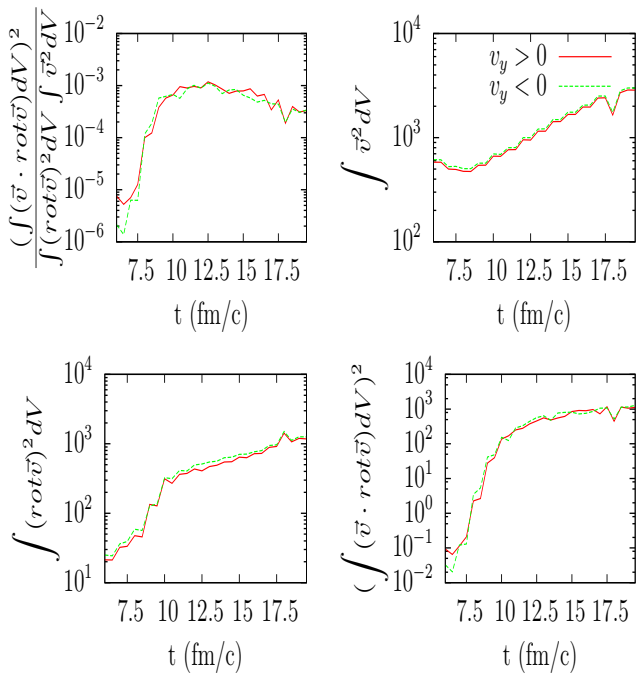

  {\fontsize{12}{16} \selectfont
  \resizebox{\hsize}{4.5cm}{\input{integrplots/ivp_ratio.tex}
    \input{integrplots/ivp_v2.tex}}
  %% \resizebox{0.5\hsize}{!}{\input{integrplots/ivp_v2.tex}}\\
  \resizebox{\hsize}{4.5cm}{\input{integrplots/ivp_rotv2.tex}
    \input{integrplots/ivp_dotprod.tex}}
  }
  %% \resizebox{0.5\hsize}{!}{\input{integrplots/ivp_dotprod.tex}}
  \caption{Integral values calculated for impact parameter $b=8fm$.}
  \label{fig:integrhsd}
\end{figure}

Some integral values have also been computed for additional information and
comparison. Plots for these values are presented in
Figure~\ref{fig:integrhsd}. Both models give very
similar results. However there is a distinct peak on the plot for
$$\frac{(\int (\vec{v}, rot \vec{v})dV)^2}{\int v^2dV \int (rot \vec{v})^2dV}$$
in QGSM model, that is missing on the corresponding plot in HSD model.
The value eventually decreases but not so rapidly. In both cases this values
is very small (~$10^{-2} - 10^{-3}$).\\

In this section we have studied hydrodynamic vorticity, helicity and some
integral values in heavy ion collisions. The results were compared to those
that were obtained with the help of QGSM model. They are mostly similar
except for an explainable shift in time. Although the integral values
are small in magnitude, their time dependence resembles the results obtained
in the QGSM model.

The possible observable result of non-zero medium helicity is polarization
of $\Lambda$ - hyperons with different signs of $y$ - component of momentum.
A quantitive estimate of such polarization is given in \cite{helsep0}. At
helicity values calculated here and in \cite{helsep0} the $\Lambda$ - hyperon
polarization is possible to observe. The possibility of $\Lambda$ - 
hyperon polarization effect is also discussed in \cite{Rogachevsky:2010ys}.
$\Lambda$ - hyperon polarization is considered in hydrodynamic model in
\cite{lambdapol}.

\section{Handedness}

Since nuclei have non-zero angular momentum in non-central collisions
we can expect to find some p-odd effects in the final state. In this part
of the article we will try to find relation between properties of
particles in the final state with parity in the initial state.

To obtain information about polarization of particles in the initial state
based on the properties of particles in the final state several
methods were proposed \cite{emth} \cite{nachtm}. These methods are based
on computation of vector or triple product of 3-momenta of particles
in the final state. These methods are suitable for processing experimental
data.

In the first article \cite{nachtm} a pseudoscalar $T$ was introduced:
$$T = \frac{1}{|\vec{p}|}([\vec{p_1}, \vec{p_2}], \vec{p_3}),$$
with $|\vec{p_1}| > |\vec{p_2}| > |\vec{p_3}|$, where $\vec{p_1}$, $\vec{p_2}$
and $\vec{p_3}$ - 3-momenta of particles in the final state, 
$\vec{p}$ - momentum of the particle in the initial state. Using $T$ and
quantities derived from it some reactions including electron-positron
annihilation to hadrons and nucleon collisions were considered.

Later, in \cite{emth} a new quantity called handedness was defined.
It was proposed to investigate polarization of the initial
quark or gluon. Longitudinal handedness is defined as follows:
$$ H_{||} = \frac{N_{l} - N_{r}}{N_{l} + N_{r}}, $$
where $N_{l}$ and $N_{r}$ - is the number of left- and right-handed
combinations $\vec{k}$, $\vec{k_1}$, $\vec{k_2}$:

$$
\begin{cases}
e_{ijk}k^ik_1^jk_2^k > 0, & \text{for  }  N_l, \\
e_{ijk}k^ik_1^jk_2^k < 0, & \text{for  }  N_r.
\end{cases}
$$
Here, $\vec{k}$ - momentum of the initial particle,
$\vec{k_1}, \vec{k_2}$ - momenta of particles (pions) in the final state.
It was proposed to sort particles $\vec{k_1}$ and $\vec{k_2}$
according to their charge or magnitudes of momenta
Two transverse-handedness parameters.

\subsection{Methods and results}

Based on these articles we can introduce the following quantity:
$$
\eta=\frac{\sum(\vec{p_3}, \vec{p_2}, \vec{p_1})}
          {\sum|(\vec{p_3}, \vec{p_2}, \vec{p_1})|},
$$
where $(\vec{p_3}, \vec{p_2}, \vec{p_1})$ - triple product
$(\vec{p_3},[\vec{p_2},\vec{p_1}])$ with all vectors in a
triplet in the same octant in the momentum space,
$\vec{p_1}, \vec{p_2}, \vec{p_3}$ - momenta of pions in the final state.
Momenta in each triple product were sorted: 
$$|p_3|^2 < |p_2|^2 < |p_1|^2.$$
Hence eight values $\eta_i, i = 0..7$, one for each octant, were
calculated. Octants were enumerated the way described in
table ~\ref{tab:octenum}.
Au+Au collisions were considered with projectile energy of 5GeV per
nucleon in the laboratory frame with impact parameter $b = 7 fm/c$.
Heavy-ion collisions were modelled,
as before, in Hadron-String Dynamics model \cite{hsdref}.

\begin{figure}[!h]
  \centering
  \resizebox{\hsize}{!}{\input{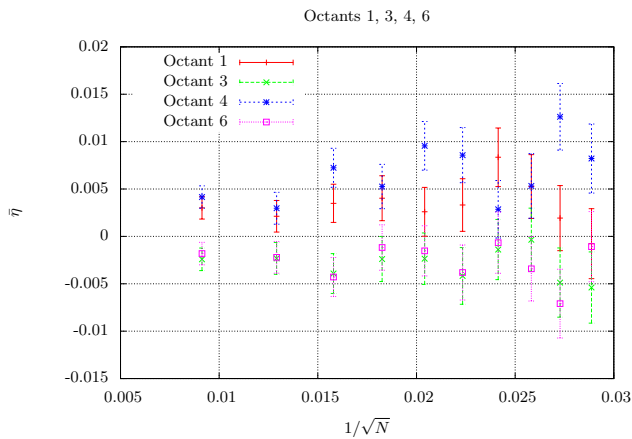}}
  \caption{Dependence of $\bar{\eta}$ on $1/\sqrt{N}$,
    for impact parameter $b=7fm$. Octants 1, 3, 4, 6.}
  \label{fig:out_oct0}
\end{figure}

\begin{figure}[!h]
  \centering
  \resizebox{\hsize}{!}{\input{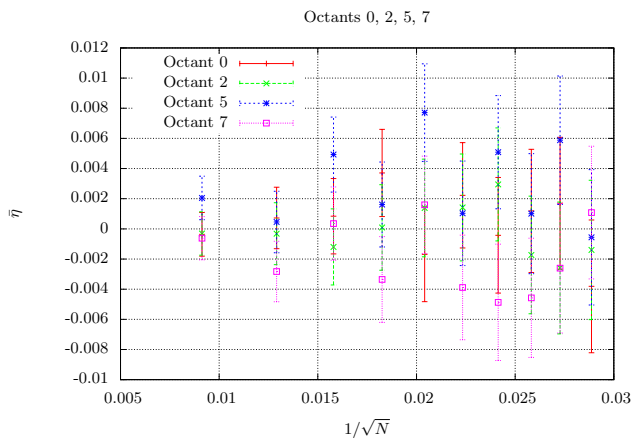}}
  \caption{Dependence of $\bar{\eta}$ on $1/\sqrt{N}$,
    for impact parameter $b=7fm$. Octants 0, 2, 5, 7}
  \label{fig:cum_out}
\end{figure}

Since collisions are non-central, non-zero values of $\eta$ are expected.
To take into account statistical errors, $\eta$ was averaged over a number
of events and an estimate of standard deviation for every average value was
taken to be the statistical error. Average $\bar{\eta}$ is plotted with
the estimate of standard deviation for every octant over $1/\sqrt{N}$, where
$N$ - is the number of events used to calculate the average value
(Figures ~\ref{fig:out_oct0} and ~\ref{fig:cum_out}.)
\begin{center}
  \begin{table}
    \begin{tabular}{|l|l|}
      \hline
      Octant & Momentum \\
      \hline
      0 & $p_x > 0, p_y > 0, p_z > 0$\\
      \hline
      1 & $p_x > 0, p_y > 0, p_z \leq 0$\\
      \hline
      2 & $p_x > 0, p_y \leq 0, p_z > 0$\\
      \hline
      3 & $p_x > 0, p_y \leq 0, p_z \leq 0$\\
      \hline
      4 & $p_x \leq 0, p_y > 0, p_z > 0$\\
      \hline
      5 & $p_x \leq 0, p_y > 0, p_z \leq 0$\\
      \hline
      6 & $p_x \leq 0, p_y \leq 0, p_z > 0$\\
      \hline
      7 & $p_x \leq 0, p_y \leq 0, p_z \leq 0$\\
      \hline
    \end{tabular}
    \caption{Octant enumeration.}
    \label{tab:octenum}
  \end{table}
\end{center}

Although the statistical error is high at low $N$, we can see that it
decreases at higher $N$. As the number of events $N$ increases, $\bar{\eta}$
in octants 1, 3, 4 and 6 does not completely vanish. Moreover $|\bar{\eta}|$
is higher than one standard deviation. This points to the possibility of
non-zero values of $\bar{\eta}$ in non-central collisions.

\section{Conclusion}

We have studied vorticity and helicity in heavy-ion collisions in the HSD
model for Au+Au reactions at small energy $\sqrt{s} = 5 GeV$ and for
different impact parameters.

Using hydrodynamic approach we calculated the velocity field of the
final state particles. Using this velocity field we calculated the averaged
weighted voritcity and studied its time evolution. We noticed
that the average weighted $y$-component of vorticity
decreases over time in non-central
heavy-ion collisions and disappears for the central collisions.
The spacial distribution averaged over all $x-z$ planes was also considered.
The difference of the emerging picture with that in the hydrodynamical
approach\cite{cshid} is due to the viscosity effects. 

Helicity separation was observed in the HSD model. The results in this model
are similar to those that were obtained in the QGSM model \cite{helsep0}, with
some differences in time dependence and in magnitude. The most significant
discrepancy - in the time dependence can be explained by details of heavy ion
collision simulation. At the initial moment of time $t = 0$ there is a
significant distance between the nuclei, so the reaction happens later.
The difference in magnitude isn't significant. Generally the integral values
have similar time dependence, but have smaller magnitude in the HSD model.
Non-zero helicity in such reactions can result in $\Lambda$ - hyperon
polarization which can be observed.

We have also proposed a pseudoscalar quantity $\eta$ for investigation
of parity-odd effects in heavy-ion collisions based on previous suggestions
\cite{nachtm} \cite{emth}. The advantage of this approach is
suitability for experimental observations without additional calculations.
Using computer simulations in the HSD model we have obtained preliminary
results for $\bar{\eta}(1/\sqrt{N})$ dependence indicating that it could be
used to probe for p-odd effects in non-central collisions. Note the
"handedness separation" to the different sides of reaction plane
similar to helicity separation discussed above.
\\\\

\centerline{\textbf{Acknowledgements}}

Authors are grateful to E.L. Bratkovskaya and M. Baznat for help,
discussions and comments. 
O.T. is  also thankful to  L.~P.~Csernai, A.V. Efremov, K. Gudima and
A.S. Sorin for stimulating discussions and valuable remarks.

\end{document}